\documentclass[aps,apl,notitlepage,twocolumn,superscriptaddress]{revtex4-1}

\usepackage{amsmath}
\usepackage{amssymb}
\usepackage{graphicx}
\usepackage{epstopdf}
\usepackage[colorlinks=true]{hyperref}

\begin{document}
\title{Nonlocal Spin Transport Mediated by a Vortex Liquid in Superconductors}

\author{Se Kwon Kim}
\affiliation{Department of Physics and Astronomy, University of California, Los Angeles, California 90095, USA}
\author{Roberto Myers}
\affiliation{Department of Materials Science and Engineering, The Ohio State University, Columbus, Ohio 43210, USA}
\affiliation{Department of Electrical and Computer Engineering, The Ohio State University, Columbus, Ohio 43210, USA}
\author{Yaroslav Tserkovnyak}
\affiliation{Department of Physics and Astronomy, University of California, Los Angeles, California 90095, USA}

\begin{abstract}
Departing from the conventional view on superconducting vortices as a parasitic source of dissipation for charge transport, we propose to use mobile vortices as topologically-stable information carriers for spin transport. To this end, we start by constructing a phenomenological theory for the interconversion between spin and vorticity, a topological charge carried by vortices, at the interface between a magnetic insulator and a superconductor, by invoking the interfacial spin Hall effect therein. We then show that a vortex liquid in superconductors can serve as a spin-transport channel between two magnetic insulators by encoding spin information in the vorticity. The vortex-mediated nonlocal signal between the two magnetic insulators is shown to decay algebraically as a function of their separation, contrasting with the exponential decay of the quasiparticle-mediated spin transport. We envision that hydrodynamics of topological excitations, such as vortices in superconductors and domain walls in magnets, may serve as a universal framework to discuss long-range transport properties of ordered materials.
\end{abstract}

\date{\today}
\maketitle

\emph{Introduction.}|Superconductivity refers to the phenomenon of a topologically-protected collective charge transport. It has been one of the central topics in physics because of practical motivations, e.g., for long-distance power transmission, as well as fundamental interest in quantum phases of matter~\cite{*[][{, and references therein.}] Tinkham2004}. A superconductor loses its ability for lossless transport when the phase coherence of the condensate wavefunction is destroyed by the proliferation of vortices, topological phase defects, which are inevitable in low-dimensional structures~\cite{*[][{, and references therein.}] BlatterRMP1994, *[][{, and references therein.}] HalperinIJMPB2010}. Since the motion of vortices gives rise to a finite resistance in superconductors, a central goal in the materials engineering of superconductors has been to reduce the number of thermal vortices and immobilize them by engineering pinning defects~\cite{*[][{, and references therein.}] KwokRPP2016}.

In contrast to the conventional, antagonistic view on vortices in superconductors for charge transport, in this Letter, we propose to use mobile superconducting vortices as efficient information carriers for spin transport, which are endowed with a stability by their topological characteristics. Analogous research has been done previously for topological solitons in magnets, such as domain walls~\cite{ParkinScience2008} and skyrmions~\cite{*[][{, and references therein.}] NagaosaNN2013}. Topological magnetic solitons can store information in their topological charges: the chirality of a domain wall and the winding number of a skyrmion. The stability associated with the topological characteristics allows them to transport information over relatively long distances, compared to quasiparticles, such as magnons with a finite lifetime, giving rise to an algebraically decaying nonlocal transport~\cite{KimPRB2015-5, OchoaPRB2016}.

\begin{figure}
\includegraphics[width=\columnwidth]{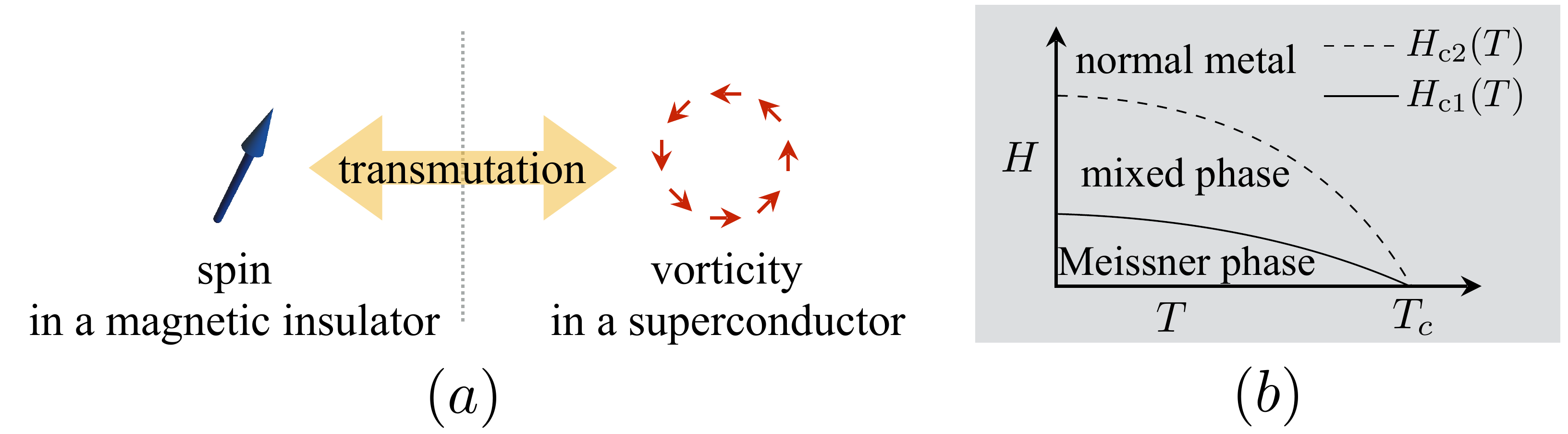}
\caption{(a) A schematic of spin-vorticity transmutation at the interface between a magnetic insulator and a superconductor. The blue arrow represents spin in a magnetic insulator and the red arrows depict the phase of a superconducting vortex. (b) A schematic of the mean-field phase diagram of bulk type-II superconductors~\cite{BlatterRMP1994}, which comprises the Meissner phase at low magnetic fields $H < H_\text{c1}$ with complete expulsion of the magnetic flux, the mixed phase at intermediate fields $H_\text{c1} < H < H_\text{c2}$, where the magnetic flux penetrates into a superconductor in the form of vortices, and the normal-metal phase at high fields $H > H_\text{c2}$, where superconductivity is destroyed. $T_c$ is the critical temperature for the onset of superconductivity. Vortices in the mixed phase can form several states of matter, including vortex liquid, which is the phase of our main interest. We remark that our theory for spin-vorticity transmutation works for any superconductor that supports vortex excitations.}
\label{fig:fig1}
\end{figure}

In this Letter, we show that topological defects in superconductors, vortices, can transport spin information efficiently by encoding it in their topological charge that is referred to as vorticity. Superconductors of our interest are type-II superconductors such as Nb or La$_{2-x}$Sr$_x$CuO$_4$ as well as type-I superconductor thin films that can support vortices as excitations~\cite{*[][{, and references therein.}] BlatterRMP1994, LasherPR1967, *SweenyPRB2010}. For example, the schematic mean-field phase diagram of bulk type-II superconductors is shown in Fig.~\ref{fig:fig1}(b)~\cite{BlatterRMP1994}. The mixed (or Schubnikov) phase harbors superconducting vortices, and they can form various states of matter such as vortex lattice or vortex liquid~\cite{AndersonNP2007, *LiNP2007, *KivelsonPhysics2010}. We will focus on the vortex-liquid phase of superconductors denoted by VL, where vortices can flow individually, in this Letter. Specifically, first, we develop a phenomenological theory for the interconversion between spin in a magnetic insulator and vorticity in a superconductor at the interface, which is schematically illustrated in Fig.~\ref{fig:fig1}(a). One process for the interconversion, which occurs via the interfacial spin Hall effect, is given for concreteness. Second, based on the aforementioned theory for spin-vorticity interconversion, we show that vortices in a superconductor can support algebraically-decaying nonlocal spin transport between two distant magnetic insulators sandwiching a superconductor. We will conclude the Letter by discussing other possible mechanisms for spin-vorticity interconversion and providing some future outlooks. We envision that the field of superconducting spintronics~\cite{LinderNP2015}, in which the interaction between a magnet and a superconductor has been explored mainly focusing on spin-polarization of quasiparticles, can be enriched by incorporating the hitherto largely ignored objects---vortices---as active ingredients along with a natural spin-vorticity transmutation. The materials library of previously considered low-performance superconductors (with mobile vortices), for example those with small lower critical field $H_{c1}$, can be revisited for their potential as efficient spin-vortex conversion layers.

\begin{figure}
\includegraphics[width=0.8\columnwidth]{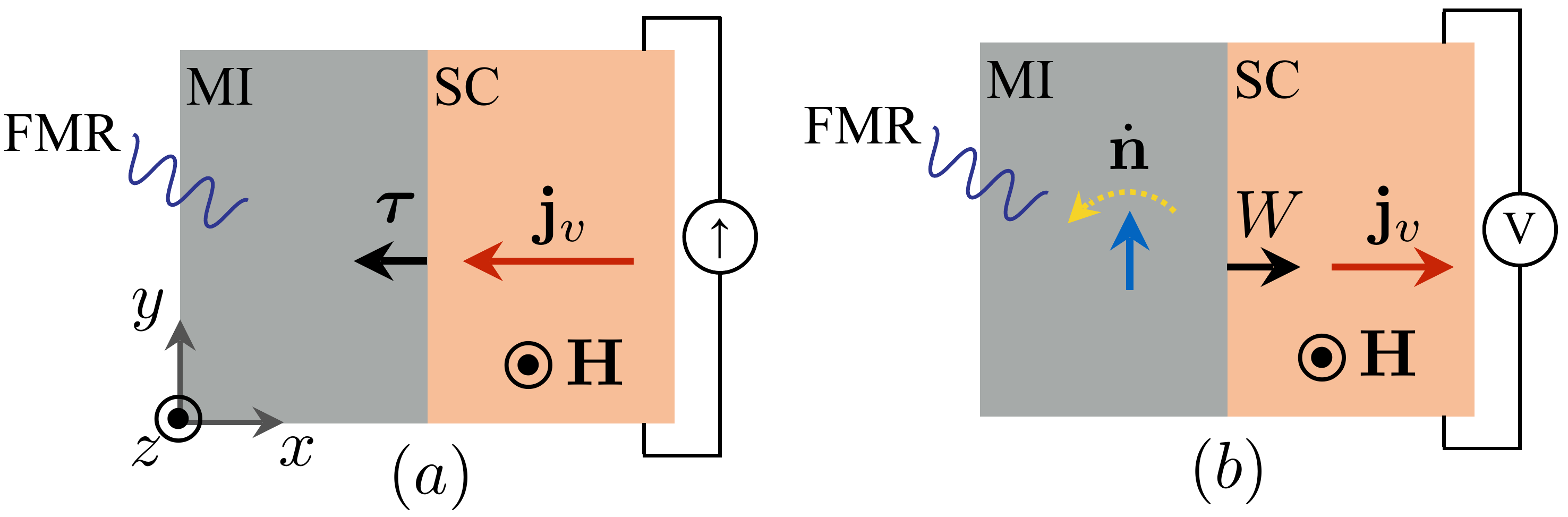}
\caption{Schematics of experimental setups for probing spin-vorticity transmutation at the interface between a magnetic insulator (MI) and a superconductor (SC) subjected to an external magnetic field $\mathbf{H}$. (a) The torque $\boldsymbol{\tau}$ [Eq.~(\ref{eq:torque})] on MI exerted by the vorticity flux $\mathbf{j}_v$ in SC can be probed by performing ferromagnetic resonance (FMR) measurements on MI, while applying a transverse current to SC, which generates a longitudinal vorticity flux $\mathbf{j}_v$ via the Lorentz force~\cite{AndersonRMP1964, *BardeenPR1965, *HuebenerPRB1969, *CampbellAP1972, *AoPRL1993, *ThoulessPRL1996, *WangPRB2006, *SoninPRB2013}. (b) The vorticity-dependent work $W$ [Eq.~(\ref{eq:work})] performed by magnetic dynamics $\dot{\mathbf{n}}$ on vortices creates the nonequilibrium vorticity accumulation at the interface and thereby induces a diffusive vorticity flux $\mathbf{j}_v$. The vorticity flux $\mathbf{j}_v$ can be probed by measuring a transverse voltage drop across the SC, which is induced by $\mathbf{j}_v$ via the Josephson effect~\cite{AndersonRMP1964}, while driving FMR dynamics in the MI. In addition, the energy dissipation through vortex generation suggests a new channel for Gilbert damping of MI, which can be manifested through the enhancement FMR linewidth~\cite{JeonNM2018}.}
\label{fig:fig2}
\end{figure}

\emph{Spin and vorticity.}|We provide a phenomenological theory for the interconversion between spin in a magnetic insulator (MI) and vorticity in a superconductor (SC) harboring a vortex liquid~\cite{AndersonNP2007}. For concreteness, we consider materials lying on the $xy$ plane that share two-dimensional interfaces in the $yz$ plane, and will assume that the magnetic order parameter and the superconducting wavefunction are uniform along the $z$ direction. See Fig.~\ref{fig:fig2} for schematics. A vortex in a superconductor is characterized by its vorticity,
\begin{equation}
q = \frac{1}{2 \pi} \oint d\mathbf{r} \cdot \boldsymbol{\nabla} \phi \, ,
\end{equation}
an integer number measuring how many times the phase $\phi$ of the condensate wavefunction winds the unit circle when moving along a closed line encircling the vortex core~\footnote{The vorticity, which is also referred to as the circulation, is associated with the magnetic flux bound to the vortex, which is quantized as $\mathbf{\Phi} = - q \Phi_0 \hat{\mathbf{z}}$ in bulk when the current is fully screened~\cite{BlatterRMP1994, Tinkham2004}. Here, the negative-sign prefactor reflects the negative electric charge of Cooper pairs. Note that the magnetic flux is an axial vector similar to spin. We remark that our theory for a spin-vorticity transmutation does not rely on the quantization of magnetic flux bound to vortices.}. We focus on the elementary vortices with the unit vorticity, $q = \pm 1$, since the other vortices are suppressed energetically (while elementary vortices with the same charge repel each other).

When the Cooper pairs are stable in a superconductor, the wavefunction is well defined and the total vorticity is conserved due to its topological nature, which allows us to use the hydrodynamic theory to describe their macroscopic dynamics. The relevant hydrodynamic variables are the vorticity density $\rho_v = \rho_+ - \rho_-$ (per unit area)~\footnote{The vorticity density $\rho_v$ is the coarse-grained version of $\hat{\mathbf{z}} \cdot \left[ \boldsymbol{\nabla} \text{Re}(\psi) \times \boldsymbol{\nabla} \text{Im}(\psi) \right] / 2 \pi |\psi_0|^2$, where $|\psi_0|$ is the order-parameter magnitude away from a vortex core.}, and the vorticity-current density $\mathbf{j}_v = \mathbf{j}_+ - \mathbf{j}_-$ (per unit length), where $\rho_q$ and $\mathbf{j}_q$ are the number density and the current density of vortices with the vorticity $q = \pm 1$. At temperature $T$, vortices are nucleated and annihilated by thermal fluctuations, giving $\rho_q^0 \propto \exp(- E_q / k_B T)$ at sufficiently low temperatures in equilibrium, where $E_q$ is the energy of a vortex with the vorticity $q$ and $k_B$ is the Boltzmann constant. In equilibrium, the vorticity current vanishes, $\mathbf{j}_v \equiv 0$, but the vorticity density can be finite if the energy of a vortex depends on the vorticity, $E_+ \neq E_-$, due to, e.g., an external magnetic field. Note that, differing from the total vorticity, the total number of vortices is not conserved in the bulk since two vortices with the opposite vorticities can be nucleated and annihilated together. 

Our main results for the spin-vorticity interconversion, which we obtain below, can be summarized as follows:
\begin{subequations}
\label{eq:main1}
\begin{align}
\boldsymbol{\tau} &= (g' + g \mathbf{n} \times) (\mathbf{n} \times j_v \hat{\mathbf{z}}) \, , \label{eq:torque} \\
W &= - q [(g' + g \mathbf{n} \times) \dot{\mathbf{n}}] \cdot \hat{\mathbf{z}} \, , \label{eq:work}
\end{align}
\end{subequations}
which are related by Onsager reciprocity. The first equation describes the spin torque (per unit length) on the magnetic material induced by the vorticity-current density $j_v$ toward the magnetic insulator in the perpendicular direction to the interface, which is equal to the annihilation rate of vortices per unit length. The phenomenological coefficients $g$ and $g'$ have the unit of the angular momentum, which quantify the amount of spin transferred to the magnetic material by annihilation of one unit of the vorticity in a dissipative and reactive fashion, respectively. The second equation describes the work $W$ performed by slow magnetic dynamics $\dot{\mathbf{n}}$ on a single nucleated vortex with the vorticity $q = \pm 1$ entering the superconductor through the interface. Schematics of experimental setups for probing our results for the spin-vorticity interconversion are shown in Figs.~\ref{fig:fig2}. We remark that materials with higher $T_c$ are better for probing our theory since it is based on thermally populated vortices.

\begin{figure}
\includegraphics[width=\columnwidth]{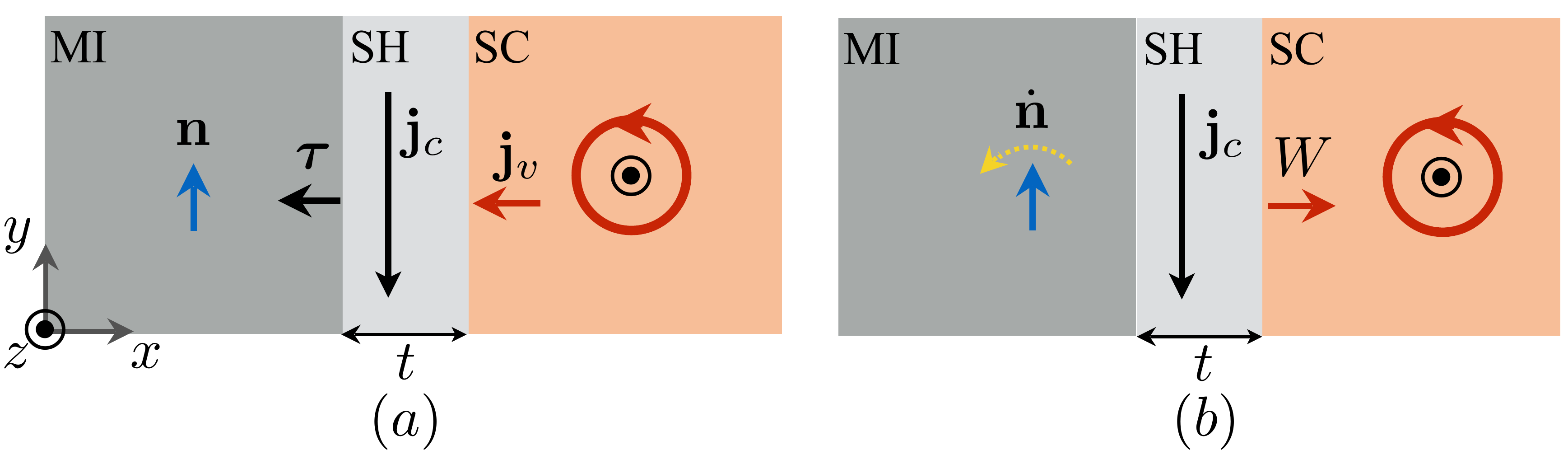}
\caption{(a) The vorticity-current density $\mathbf{j}_v$ induces the charge-current density $\mathbf{j}_c$ [Eq.~(\ref{eq:jc1})] in SH via the Josephson effect, which in turn exerts the torque $\boldsymbol{\tau}$ [Eq.~(\ref{eq:tau})] on the MI via the spin Hall effect. (b) The dynamics of the MI $\dot{\mathbf{n}}$ induces the normal charge-current density $\mathbf{j}_c$ [Eq.~(\ref{eq:jc2})] in SH via the inverse spin Hall effect. The counter-propagating supercurrent then performs the vorticity-dependent work $W$ [Eq.~(\ref{eq:W})] on a vortex via the Lorentz force.}
\label{fig:fig3}
\end{figure}

One concrete toy model for the above results, which will be given below, connects spin in a magnetic insulator and vorticity in a superconductor via charge at the interface. The central step to understand the processes is to recognize that a region (denoted by SH in Fig.~\ref{fig:fig3}) of the superconductor interfacing with the magnetic insulator will be subject to the interfacial spin Hall effects~\cite{TserkovnyakPRB2014, EmoriPRB2016}, by which a normal charge current can induce the spin-transfer torque on the magnetic insulator and, reciprocally, the magnetic dynamics can induce the charge current in the superconductor. The effective thickness of the region SH will be denoted by $t$. Here, we would like to emphasize that this is just a toy model, in which the region SH is introduced conceptually to connect spin and vorticity via charge with separate accounts of spin-charge coupling and charge-vortex coupling. The phenomenology in Eqs.~(\ref{eq:main1}) should work generally, subject to any spin-orbit coupling at the interface.

Let us first describe the process for the torque on the magnetic insulator induced by the vorticity current in the superconductor. See Fig.~\ref{fig:fig3}(a) for the schematic geometry. First, the vorticity-current density $\mathbf{j}_v = - j_v \hat{\mathbf{x}}$ induces the transverse electric field, $\mathbf{E} = \mathbf{j}_v \times \Phi_0 \hat{\mathbf{z}}$~\cite{AndersonRMP1964}, which can be understood as a manifestation of the Josephson relation via the vortex flow~\footnote{Alternatively, it can be understood as a manifestation of Faraday's law: a moving magnetic flux induces a transverse electric field.}. Here, $\Phi_0 = h / 2 e$ is the magnetic flux quantum, in terms of the Planck constant $h$ and the magnitude of the electron charge $e > 0$. The induced diffusive charge-current density carried by the normal component in the region SH~\footnote{At steady state, there will be no net charge current in the SH region because the diffusive charge-current carried by the normal component will be counterbalanced by the opposite charge current carried by the superconducting component. Within our treatment of spin-orbit coupling at the interface, the diffusive charge current induces the reactive and the dissipative torques on the magnetic insulator, whereas the supercurrent does not induce any. The supercurrent, however, might induce a reactive torque via other mechanisms.} is then given by
\begin{equation}
\label{eq:jc1}
\mathbf{j}_c = \sigma \mathbf{E} = \sigma \Phi_0 j_v \hat{\mathbf{y}} \, .
\end{equation}
Here, $\sigma \Phi_0$ has the unit of electric charge, parametrizing the interconversion efficiency from the vorticity current to the charge current. Due to the spin Hall effects, the charge current parallel to the interface gives rise to the torque on the magnetic insulator, which can be written as
\begin{equation}
\label{eq:tau}
\boldsymbol{\tau} = (\sigma \Phi_0 t) (\eta + \vartheta \mathbf{n} \times) (\mathbf{n} \times j_v \hat{\mathbf{z}}) \, ,
\end{equation}
within the spin Hall phenomenology~\cite{TserkovnyakPRB2014}, where $\eta$ and $\vartheta$ quantify the reactive and dissipative torques, respectively. Here, $j_v$ is the rate of annihilation of vortices at the interface MI$|$SH per unit length. By comparing the obtained expression to the first equation in the main results [Eq.~(\ref{eq:main1})], we can identify the coefficients: $g = \sigma \Phi_0 t \vartheta$ and $g' = \sigma \Phi_0 t \eta$ are the spin angular momentum transferred to the magnetic insulator during the annihilation of one vorticity via the dissipative and the reactive processes, respectively. The dissipative coefficient $\vartheta$ can be written as $ \vartheta = (\hbar / 2 e t) \tan \Theta$, with $\Theta$ identified as the effective spin Hall angle in the SH region. If we use the material parameters of platinum, $\sigma \sim 10^7$ ($\Omega$m)$^{-1}$ and $\Theta \sim 0.1$~\cite{LiuarXiv2011}, we obtain $g \sim 600 \, \hbar$; the annihilation of a single vortex can pump hundreds of spins (in units of $\hbar$).

Next, let us turn to the reciprocal process for the work on a vortex done by the magnetic dynamics. The dynamics of the magnetic insulator gives rise to the electromotive force in the region SH, $\boldsymbol{\epsilon} = - [(\eta + \vartheta \mathbf{n} \times) \dot{\mathbf{n}}] \times \hat{\mathbf{x}}$, via the spin Hall effects~\cite{TserkovnyakPRB2014}. Here, $\hat{\mathbf{x}}$ is the unit vector perpendicular to the interface plane. It subsequently induces the diffusive charge-current density in the region SH, 
\begin{equation}
\label{eq:jc2}
\mathbf{j}_c = - \sigma \left\{ [(\eta + \vartheta \mathbf{n} \times) \dot{\mathbf{n}}] \cdot \hat{\mathbf{z}} \right\} \hat{\mathbf{y}} \, .
\end{equation}
In the steady state, where there must be no net charge flow, this diffusive current is counterbalanced by the super current, $\mathbf{I}_s = - t \mathbf{j}_c$.

The induced super current $\mathbf{I}_s$ then exerts the transverse Lorentz force on a vortex~\cite{AndersonRMP1964}, which can be related to the Josephson relation mentioned above or, equivalently, Faraday's law by invoking Onsager reciprocity. The work performed by the Lorentz force on a vortex with vorticity $q$ is given by
\begin{equation}
\label{eq:W}
W = - q \left( \mathbf{I}_s \times \Phi_0 \hat{\mathbf{z}} \right) \cdot \hat{\mathbf{x}} = - q (\sigma \Phi_0 t) [(\eta + \vartheta \mathbf{n} \times) \dot{\mathbf{n}}] \cdot \hat{\mathbf{z}} \, ,
\end{equation}
yielding Eq.~(\ref{eq:work}), as is expected from Onsager reciprocity.

\begin{figure}
\includegraphics[width=0.8\columnwidth]{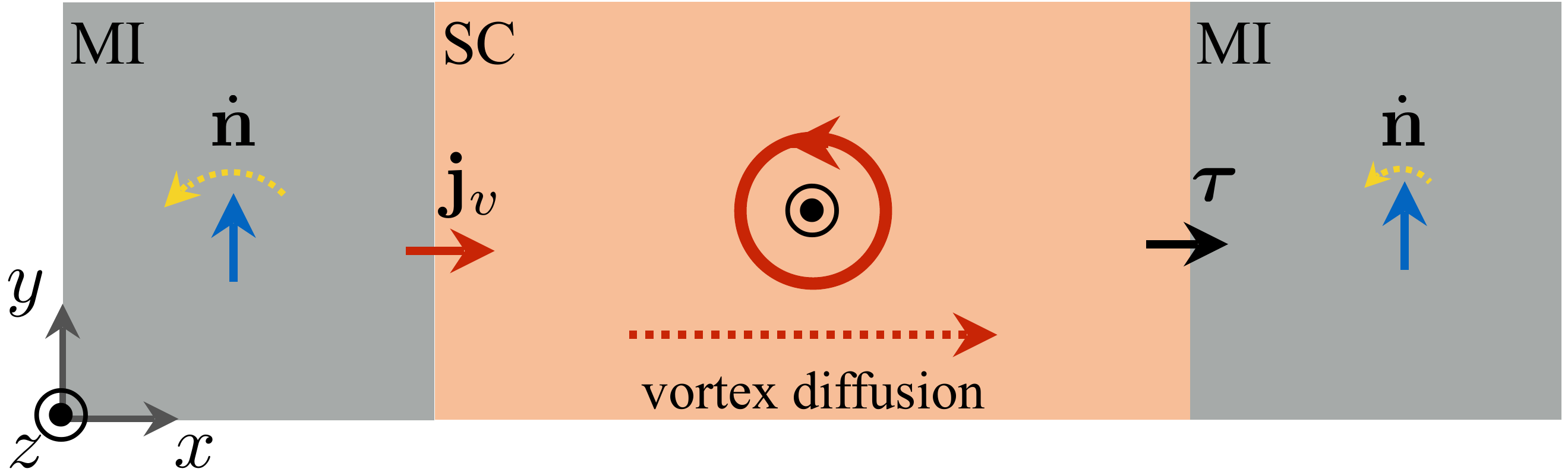}
\caption{A schematic of the nonlocal spin transport between two MIs mediated by a vortex liquid in the interconnecting SC. The dynamics of the magnetization $\mathbf{n}$ of the left MI pumps the vorticity current $\mathbf{j}_v$ into the SC. Some of the pumped vortices travel across the SC by thermal diffusion and reach the interface to the right MI, exerting the torque $\boldsymbol{\tau}$ to it. The width of the SC needs to be much larger than the length to prevent vortices from escaping through sides.}
\label{fig:fig4}
\end{figure}

\emph{Nonlocal spin transport by a vortex liquid.}|The vorticity pumping by the magnetic dynamics and the reciprocal torque by the vorticity flux can be used to transport spin between two distant magnetic insulators via a vortex liquid in type-II superconductors, which we shall study below based on our results in Eqs.~(\ref{eq:main1}). The geometry that we consider is schematically drawn in Fig.~\ref{fig:fig4}. 

The left MI is a ferromagnet at resonance with a rf field under a magnetic field in the $z$ direction, which serves as the spin battery~\cite{BrataasPRB2002} in our setup. The order parameter $\mathbf{n}$ of the left MI precesses with the cone angle $\theta$ around the $z$ axis at frequency $\omega$. According to Eq.~(\ref{eq:work}), the magnetic order-parameter dynamics performs the vorticity-dependent work, $W_{\pm} = \mp g \omega \sin^2 \theta$, on a superconducting vortex when it enters the superconductor. Here, the reactive part $\propto g'$ vanishes after being averaged over time. This vorticity-dependent work pumps the vorticity into the superconductor through the interface according to the reaction-rate theory as follows~\cite{HanggiRMP1990, BartolfPRB2010, *NastiPRB2015}. The nucleation rate for a vortex with vorticity $q = \pm 1$ per unit length along the $y$ direction is described by $\Gamma_\pm = \nu \exp(- E_\pm / k_B T)$, where $\nu$ is the attempt frequency and $E_\pm$ is the energy barrier for the nucleation of a vortex. The annihilation rate per unit length is described by $\gamma \rho_\pm$, where $\gamma$ is the annihilation rate per unit density that does not depend on the vorticity. In equilibrium, the nucleation and the annihilation rates must be equal, giving $\Gamma_\pm = \gamma \rho_\pm$, forming the balance equations.

The dynamics of the adjacent magnetic insulator breaks the balance equations as follows. The work done by the magnetic dynamics of the left MI modifies the energy barrier, $E_\pm = E_\pm^0 - W_\pm$, from its equilibrium value $E_\pm^0$ by which the nucleation rate changes as well, $\Gamma_\pm = \Gamma^0_\pm (1 \mp g \omega \sin^2 \theta / k_B T)$ in linear order. Then, the net injection rate of the vorticity (per unit length) is given by $- (\Gamma^0_+ + \Gamma^0_-) g \omega \sin^2 \theta / k_B T = - \gamma \rho^0 g \omega \sin^2 \theta / k_B T$. Therefore, the vorticity-current density at the interface of MI$|$SC is given by
\begin{equation}
\label{eq:left}
j_v (x = 0) = - \gamma \rho^0 g \omega \sin^2 \theta / k_B T - \gamma \delta \rho_v (x = 0) \, ,
\end{equation}
where $\delta \rho_v \equiv \rho_v - \rho^0_v$ is the nonequilibrium vorticity density. Here and after, $j_v$ represents the component of the vorticity-current density in the $x$ direction, normal to the interfaces between MIs and SC. The first term on the right-hand side is the vorticity pumped by the magnetic dynamics; the second is the annihilation rate of the nonequilibrium vorticity density. The pumped vorticity diffuses through the bulk of SC, satisfing the continuity equation: $\partial_t \delta \rho_v + \partial_x j_v = 0$, which is rooted in the conservation of the topological charge, the net vorticity. We assume that the dynamics of vorticity is purely diffusive:
\begin{equation}
\label{eq:bulk}
j_v = - D \partial_x \delta \rho_v \, ,
\end{equation}
where $D$ is the diffusion coefficient of a vortex. The vorticity-current density at the interface between SC and the right MI is given by
\begin{equation}
\label{eq:right}
j_v (x = L) = \gamma \delta \rho_v (x = L) \, .
\end{equation}
In the steady state, the vorticity density is constant and the vorticity-current density is uniform. By solving Eqs.~(\ref{eq:left}), (\ref{eq:bulk}), and (\ref{eq:right}) for the uniform $j_v$, we obtain
\begin{equation}
j_v = - \frac{g \omega \sin^2 \theta}{2 + \gamma L / D} \frac{\rho^0}{k_B T} \, .
\end{equation}
When SC is sufficiently short, $L \ll D / \gamma$, about half of the pumped vorticity $- \gamma \rho^0 g \omega \sin^2 \theta / k_B T$ by the dynamics of the left MI passes through the superconductor and leaves the superconductor through the interface to the right MI. The vorticity annihilation at the interface SC$|$MI exerts the torque on the right MI, which can be obtained from Eq.~(\ref{eq:torque}). The resultant antidamping torque can induce the dynamics of the right MI, e.g., by driving it into an auto-oscillation phase~\cite{ChenPIEEE2016}. The modulation of the torque can also be detected magnetoresistively by a lock-in method.

\emph{Discussion.}|In this Letter, we have focused on one toy-model process of the spin-vorticity interconversion, which occurs via the interfacial spin Hall effects between a magnetic insulator and a superconductor. There can be other mechanisms that can give rise to the interconversion. For example, a vortex in a superconductor can harbor spin-polarized normal quasiparticles in its core, which tend to align along the magnetic flux associated with the vortex via the Zeeman coupling as demonstrated in the vortex-flipping experiments~\cite{PatinoPRB2013}. The spin angular momentum of quasiparticles in the vortex core can be directly transferred to a magnetic insulator when the vortex is annihilating at its interface. In addition, although the present work is focused on the vortex-liquid phase of type-II superconductors, it can be extended to the other phases such as the vortex-solid phase as done for the previous work on nonlocal information transport by the elastic response of magnetic skyrmion crystals~\cite{OchoaPRB2017}. We remark here that spin-vorticity interconversion has been studied also for mechanical rotations of fluids~\cite{TakahashiNP2015, *DoornenbalarXiv2018}.

In 1979, \textcite{MerminRMP1979} presented the theory for the classification of topological solitons in ordered media, which is determined by the topology of the order-parameter space independent of the microscopic details of the systems. Vortices in superconductors, domain walls in easy-axis magnet, and skyrmions in chiral magnet are a few examples of these topological solitons, which can be considered as emergent particles that are robust owing to their topological characteristics. They can form various phases of matter similar to elementary particles, as exemplified by vortex liquid and vortex solid phases of type-II superconductors~\cite{Blatter2008}. Our present work on long-range information transport carried by superconducting vortices along with the previous works on magnetic domain walls~\cite{KimPRB2015-5} and chiral skyrmions~\cite{OchoaPRB2016, OchoaPRB2017} leads us to envision that we may construct the theory for the hydrodynamics of topological solitons solely based on the topology of the order-parameter space, upon which the static and dynamic properties of emergent solitonic matters can be discussed generally.

\begin{acknowledgments}
We thank Chiara Ciccarelli for the enlightening discussions. This work was supported by the Army Research Office under Contract No. W911NF-14-1-0016 and the NSF-funded MRSEC under Grant No. DMR-1420451.
\end{acknowledgments}

\bibliography{/Users/evol/Dropbox/School/Research/master}

\end{document}